\documentclass[conference]{IEEEtran}

\usepackage{cite}

 \usepackage{multirow}
 \usepackage{colortbl}
 \usepackage{hhline}

\usepackage{caption}
\usepackage{subcaption}

\usepackage{algorithm}
\usepackage{algorithmic}

\usepackage{tikz}
\usetikzlibrary{shapes.geometric, arrows}

\tikzstyle{process} = [rectangle, rounded corners, minimum width=3cm, minimum height=1cm,text centered, draw=black, fill=white]
\tikzstyle{arrow} = [thick,->,>=stealth]

\usepackage{color,soul}
\usepackage{gensymb}
\usepackage{dsfont}

\ifCLASSINFOpdf
  \usepackage{graphicx}
  \graphicspath{{../pdf/}{../jpeg/}}
  \DeclareGraphicsExtensions{.pdf,.jpeg,.png}
\else
  \usepackage[dvips]{graphicx}
  \graphicspath{{../eps/}}
  \DeclareGraphicsExtensions{.eps}
\fi

\usepackage[cmex10]{amsmath}
\usepackage{amssymb}

\usepackage{euscript}

\usepackage{multirow}
\usepackage{algorithm}
\usepackage{diagbox}
\usepackage{physics}
\usepackage{tikz}

\usepackage{amsthm}
\usepackage{bm}

\usepackage{url}

\IEEEoverridecommandlockouts
\begin{document}

\title{Research on integrated intelligent energy management system based on
big data analysis and machine learning\\
}

\author{\IEEEauthorblockN{ Jinzhou~Xu, Yadan~Zhang, Paola Tapia
\IEEEauthorblockA{{Metropolitan University of Monterrey}\\
Monterrey, NL, Mexico \\
}
}

}
\maketitle

\begin{abstract}

The application of big data is one of the significant features of integrated smart energy. Applying it to the file management of integrated smart energy projects is of great significance for improving the efficiency of project management and control. This article first discussed the benefits and challenges of implementing big data analysis in document management and control of integrated smart energy projects. In addition, an implementation framework for big data analysis in integrated smart energy project document management was developed, and a method for optimizing the efficiency of integrated smart energy project document management through machine learning was proposed. Using various types of data and information generated during the project document management process, the efficiency of the entire process project document control through three different machine learning methods was optimized. The result of fitting a penalty linear regression model shows that when there is enough data as a training set, the accuracy of the model achieved can reach over 95\%. By using big data analysis and machine learning to analyze the efficiency of comprehensive smart energy project document management, it is possible to track the entire process of comprehensive smart energy project documents and optimize business processes, thereby strengthening project construction control and improving project construction efficiency.

\end{abstract}


\section{Introduction}

As the demand for energy in modern society continues to increase, power generation enterprises are facing more challenges\cite{1}. To improve efficiency and reduce costs, many power generation companies have begun to use machine learning and big data analysis technologies \cite{he2024lidar}. These technologies can help enterprises better understand their business data, thereby improving project management efficiency and decision-making quality.

Big data analysis is a technology that extracts useful information from large amounts of data. By analyzing large volumes of data, enterprises can better understand their business data and make more informed decisions \cite{mo2024make}. Machine learning, on the other hand, is an artificial intelligence technology that helps systems automatically identify patterns and make predictions from data through learning \cite{dai2024cloud, liu2024spam, li2022automated,  li2023stock}.

By integrating big data analysis and machine learning technologies, power generation enterprises can achieve comprehensive smart energy project management \cite{mo2024password, mo2024large,2}. For example, big data analysis technology can be used to monitor and predict power demand, thereby better planning energy production\cite{3}. At the same time, machine learning technology can help enterprises identify equipment failure patterns and predict and solve problems in advance, thus reducing maintenance costs \cite{zhou2024optimizing}. In addition, through comprehensive smart energy project management, power generation enterprises can also improve decision-making efficiency and quality\cite{lyu2023attention,lyu2022multimodal}. For instance, big data analysis and machine learning technologies can be used to assess project risks and opportunities, thereby making better project investment decisions \cite{unknown1,unknown2,4,5}. In the future, as technology continues to develop, comprehensive smart energy project management will play a more significant role\cite{lyu2022study}. It is expected that more technologies will be applied to energy project management, bringing more efficiency and cost advantages.

Big data analysis and machine learning-based comprehensive smart energy project document management is an effective solution that can help power generation enterprises improve the management efficiency of project documents, reduce project control costs, and enhance the informatization and digitalization level of enterprise operation and management, thereby achieving lean management\cite{cao2017structurally}. In the future, this field will continue to develop, bringing more improvements and benefits to power generation enterprises.

This article first introduces the benefits and challenges of implementing big data analysis in comprehensive smart energy project document control. An implementation framework for big data analysis in comprehensive smart energy project document management is developed, and a method for optimizing the efficiency of comprehensive smart energy project document management through machine learning is proposed. This method can achieve full-process tracking and business process optimization of comprehensive smart energy project documents, thereby strengthening project construction control and improving project construction efficiency.

\section{Application of big data technology in the energy field}\label{Sec:ProblemDescription}

Power generation enterprises have six goals: enabling users to effectively participate, adapting to all forms of power generation and storage, providing new products and services, ensuring an adequate level of power quality to meet diverse demands, optimizing asset utilization and improving efficiency, and responding quickly to disturbances and emergencies. Big data technology can be used to achieve these goals. This technology combines block computation with traditional clustering and does not conflict with traditional preprocessing methods. There are four key categories of big data technology used in the energy sector: data collection and storage, data correlation analysis, multi-source data control, and data visualization. Literature \cite{lin2020touch} and \cite{jiang2021recurrent} propose the overall architecture of a big data platform with neural networks, which can carry out big data application analysis for typical scenarios of macro energy management and provide a reference for energy management decisions. 


Power generation enterprises have two main types of renewable energy data: geospatial data and time data. Geospatial data is related to location, while time data concerns the temporal characteristics of data. For renewable energy, geospatial data may include the locations of transmission infrastructure, cities, factories, hospitals, schools, roads, etc. This data is mainly obtained using Geographic Information System (GIS) tools. Time data can include consumption patterns related to time (yearly, monthly, weekly, daily, and hourly), and energy at different times of the day or year (such as sunlight). For small areas, time data can be obtained through traditional IT systems such as SCADA. In the era of the Internet of Things, smart buildings can provide their own data through smart meters and sensors. In the absence of precise measurements or sensors, there are simplified methods to calculate this data. For example, considering the energy consumption peaks in city centers and decays towards the suburbs, given a radial profile $r$ that can be converted to Cartesian coordinates $(x, y)$, then $r^2 = x^2 + y^2$. If the city center is represented as $r = 0$, and the radial load density peaks at point $r_m$, the load distribution expression is:
\begin{equation}
Q'(r,t) = \sum_{n=1}^{N} P_{n}^{l} e^{-a_{n}(r-r_{m})^{2}} \beta_{n}(t)
\end{equation}
Here, \( Q \) is the spatial load; \( l \) is the energy type; \( n \) is the load component; \( P \) is the maximum power density of a specific load; \( \alpha \) is the width parameter; and \( \beta \) is the normalized function of the time variation of the load component.

Integrating equation (1) over the area and time intervals can yield the total load. 

The third type of energy classification data for power generation enterprises can be supply-demand classification, where users can be categorized not only by geographic region but also by their social class. Weather data (such as the angle of sunlight, wind speed and direction, temperature, pressure, cloud cover, humidity, etc.) plays an important role in the decision-making process of power generation enterprises. Therefore, the integration of supply-demand data, spatial data, and time data can support strategic decisions such as site selection for renewable energy power generation enterprises to improve output, productivity, and efficiency.

With the rapid development of big data technology, power generation enterprises have also started to apply big data technology in the field of project document management. Traditional project document management methods have many problems, such as scattered data, non-uniform documents, and difficulty in tracing. Big data technology can effectively solve these problems. Currently, the application of big data technology in the field of project document management for power generation enterprises is mainly reflected in the following aspects: data integration and management, data analysis and mining, data sharing and collaboration, data tracing and supervision. In conclusion, the application of big data technology in the field of project document management for power generation enterprises is becoming increasingly mature. With the continuous advancement of computer technology, big data technology will play an increasingly important role in the management of power generation enterprises, thereby improving the competitiveness and innovation capabilities of power generation enterprises.

\section{Big Data Technology Framework for Renewable Energy Generation}\label{Sec:FADMMVVC}

Data analysis and decision-making can be supported by processing large amounts of stored data. This data includes consumption rates, usage patterns, maintenance schedules and reports, financial data, etc. Advances in modern communication technology make the transmission of real-time data and the management of demand/supply balance possible. Sometimes traditional IT systems cannot detect power system oscillations, which is very important when using renewable energy, as these energies are likely to cause unpredictable stress on long-distance transmission lines connecting the grid to remote areas. Proper management using big data technology can facilitate demand response for grids, electric vehicles, and distributed energy. Therefore, big data can provide better and more secure two-way communication between different nodes to facilitate energy resources in the energy market [12].

Forecasting future demands for power generation enterprises, particularly in smart grids, is a very worthwhile research topic. Power generation enterprises use big data analytics to estimate several parameters that support the decision-making process, such as load planning. This is particularly important for power generation enterprises to assess the availability of energy and the ability of the grid to transmit energy. In addition, big data analytics has other important applications, including calculating equipment downtime and evaluating and analyzing system failures. By using big data analytics technology, the efficiency and robustness of generation and distribution functions can be improved. Practical tools for handling big data streams and related operations in power systems have already been developed, such as Hadoop, Apache Drill, and Storm [13-14]. Through these technologies, users can interact with machines in real-time.

\subsection{Big data technology framework}
The big data framework consists of three layers. The upper layer is dedicated to data storage, data access, and computation. The middle layer is responsible for data management and sharing, integrating data between different applications and regions; data privacy is a key issue at this layer. At the bottom layer, the data mining platform performs data preprocessing through data fusion technology. This big data structure can implement various functions, such as fault event analysis, risk analysis, prediction, maintenance management, and asset condition monitoring and evaluation, as well as air conditioning systems in buildings.

To serve the above framework, an integrated architecture based on big data analytics and cloud computing is proposed. The key components of this architecture are smart grids, big data tools, databases, and cloud environments. Big data tools are used for managing data storage and retrieval, as well as distributed storage in racks. The database stores data on customer consumption patterns, historical data on supply, demand, faults, etc.

Data security is a critical issue in big data analytics, so it is essential to ensure that the distributed energy routing process is protected from possible false data attacks. Different types of attacks may arise to affect the link status of energy supply, energy response, or energy transmission. Such false data can lead to supply-demand imbalances, indirect cost increases, and energy shortages. Suppose there is a group (number N) of customers in a region. \( DT_i \) represents the true demand of the \( i \)th customer, where \( i \in N \), and \( DF_i \) represents the false demand for the same customer injected into the system by hackers or competitors. False energy request messages are then sent to the energy demand nodes in the grid. When the grid is capable of meeting this demand, users receive more energy than they actually need. This results in a substantial loss of energy supplied to user \( i \), which can be expressed as

\begin{equation}
\Delta D_i = D_i^F - D_i^T \quad 
\end{equation}

If the demands of several customers are hacked, the total loss of energy supply in the grid can be expressed as:

\begin{equation}
\Delta D = \sum_{i \in N^F} \Delta D_i \quad 
\end{equation}

where \( N^F \) is the set of customers with false demands.

Suppose the grid supply can meet all the requested demands, even with the extra false demands. However, when the grid supply is insufficient to provide this part of the energy, some users in the grid will suffer from the consequences of power shortages.

\subsection{Technology supports project document management}

The technical architecture supports project document management in several aspects: data collection, data storage and management, data processing and analysis, data visualization, and data security.

\begin{enumerate}
    \item \textbf{Data Collection}: The big data technical architecture of renewable energy generation enterprises can collect a large amount of data generated from various power generation sites, such as wind speed, temperature, sunlight, etc. This data is of great significance for project document management and can provide reliable data basis for project implementation.

    \item \textbf{Data Storage and Management}: The big data technical architecture can store and manage the collected data, providing good data support for project documents. By adopting distributed storage and management methods, the efficiency and reliability of data storage and management are greatly improved.

    \item \textbf{Data Processing and Analysis}: Through the big data technical architecture, renewable energy generation enterprises can process and analyze the collected data, providing a large amount of data analysis and mining support for project document management. For example, data processing and analysis can be used to predict the operating conditions of wind turbines, optimize power loads, etc.

    \item \textbf{Data Visualization}: The big data technical architecture can visualize the data, providing intuitive data display, helping project document managers better understand and grasp the data, and make better decisions. For example, data visualization can intuitively display the power output of wind turbines.

    \item \textbf{Data Security}: The big data technical architecture can provide data security measures, offering important guarantees for project document management. For example, using data backup and disaster recovery technologies ensures the reliability and security of data.
\end{enumerate}

In summary, the big data technical architecture of renewable energy generation enterprises is very important for supporting project document management, providing support in data collection, storage and management, data processing and analysis, data visualization, and data security, thus offering strong data support for project document management.

\section{Smart Energy Project Management Efficiency Optimization}\label{Sec:Case}

A smart grid system requires information about consumer demand and energy supply, as well as an estimate of grid stability, to create new pricing for the sustainability of each energy unit in the smart grid system. The aim of this paper's research is to forecast and analyze changes in energy production and consumption relative to energy prices in a decentralized smart grid system by performing big data analysis on large datasets, and to optimize the efficiency of integrated smart energy project document management. To determine the grid stability of a distributed smart grid system, a mathematical model of a four-node star architecture is proposed, where 1 energy supply node serves 3 consumption nodes, as shown in Fig. 1. This model considers three input features: total power balance, energy price elasticity, and response time to price changes.

\begin{figure}[h]
\centering
\includegraphics[width=0.5\textwidth]{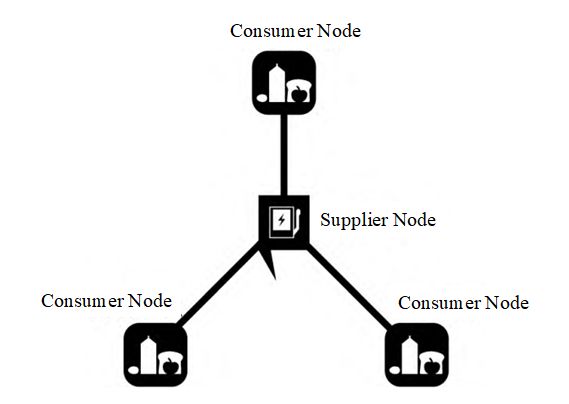}
\caption{4-node star architecture description diagram in smart grid system}
\label{fig:star_architecture}
\end{figure}

The stability of decentralized smart grid control systems can be predicted through machine learning and deep learning. This data includes information on demand inputs and grid outputs of distributed smart grid control systems collected from various resources. The steps for predicting the stability of smart grid systems using big data analytics are shown in Figure 2.

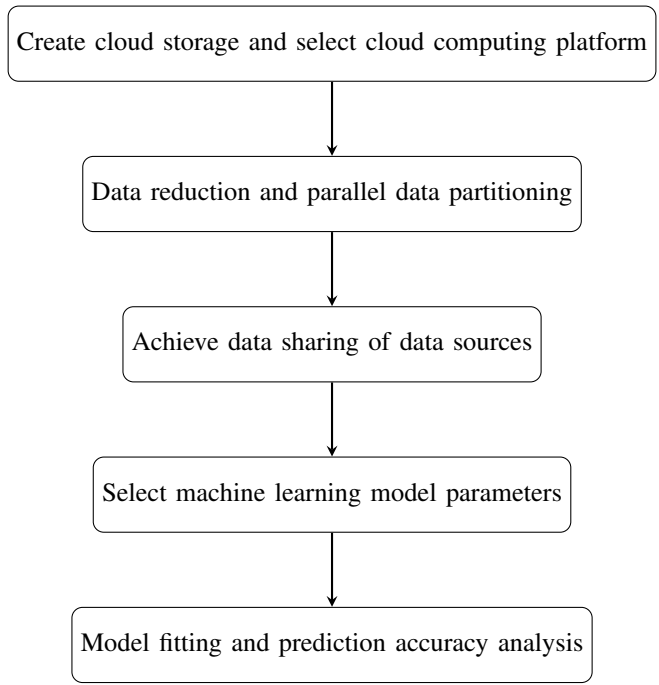
\begin{figure}[h]
    \centering
    \begin{tikzpicture}[node distance=2cm]
        \node (step1) [process] {Create cloud storage and select cloud computing platform};
        \node (step2) [process, below of=step1] {Data reduction and parallel data partitioning};
        \node (step3) [process, below of=step2] {Achieve data sharing of data sources};
        \node (step4) [process, below of=step3] {Select machine learning model parameters};
        \node (step5) [process, below of=step4] {Model fitting and prediction accuracy analysis};

        \draw [arrow] (step1) -- (step2);
        \draw [arrow] (step2) -- (step3);
        \draw [arrow] (step3) -- (step4);
        \draw [arrow] (step4) -- (step5);
    \end{tikzpicture}
    \caption{Steps for predicting big data analysis of smart grid stability}
    \label{fig:flowchart}
\end{figure}

\subsection{Create cloud storage and choose a computing platform}

Cloud storage and cloud computing are considered key steps in big data analysis and computation. Available cloud storage platforms for big data analysis include Amazon S3, BigQuery, Google Drive, Microsoft Azure, and Hadoop. In our study, Google Colab and Google Drive were used as the cloud computing platform to predict the stability of smart grids using Python coding and Apache Spark. The big data analysis was implemented in Python 3.0 on Google Collaborative, using Pyspark 3.1.2 to create data pipelines. The computations were performed on a Lenovo Intel\textregistered{} Core i5 2.70 GHz with 12 GB RAM running Windows 10 Professional operating system.

\subsection{Graph Reduction/Parallel Data Splitting}

Graph reduction data segmentation is an important step in processing big data. The purpose of graph reduction is to reduce the amount of data that needs to be processed in parallel computing and to improve data processing speed by dividing the data into homogeneous parts using parallel data segmentation. In the case study of this paper, Apache Spark was successfully used to segment the data and evaluate the accuracy of predictions using penalized regression models for each data segment.

\subsection{Implementing data contention for data sources}

The classification model in this paper was developed using supervised machine learning, where the output of the stability sequence needs to be defined as integers rather than using characters or strings. Therefore, data contention is crucial for big data. Data contention can be achieved using Pandas DataFrame or SparkSQL from the ApacheSpark library. A histogram of stability values and stability classifications of various connections in the smart grid was plotted, including positive and negative values in the "stab" column (continuous output), thus using regression. Additionally, another output is provided under the "staff" column (discrete output): if the stability index is positive, it is labeled as "stable"; if the stability index is negative, it is labeled as "unstable". Some studies were conducted, as shown in Figure 3. This model includes both discrete and continuous types, where stability is provided by the "stab" column (continuous output) for the stability index score, thus using regression. Additionally, another output is provided under the "staff" column (discrete output): if the stability index is positive, it is labeled as "stable"; if the stability index is negative, it is labeled as "unstable".

\begin{figure}[h]
\centering
\includegraphics[width=0.4\textwidth]{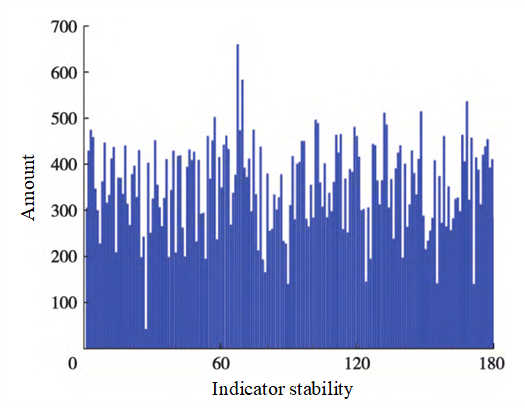}
\caption{Histogram of stability index values for the data exploration}
\label{fig:stability_histogram}
\end{figure}

\section{Simulation and result analysis}\label{Sec:Conclusion}

\subsection{Choosing machine learning model parameters}

To optimize the efficiency of integrated smart energy project document management, the dataset was collected from the smart grid system. The list of input and output of the dataset is as follows:

\begin{itemize}
    \item \textbf{tau1—tau4}: Reaction time of each network participant, with actual values ranging from 0.5 to 10.0 s (tau1 corresponds to the supplier node, tau2 to tau4 correspond to consumer nodes).
    \item \textbf{p1—p4}: Nominal power generated (positive) or consumed (negative) by each network participant, with actual values for consumers ranging from -2.0 to -0.5 (p2 to p4). Since the total consumed power equals the total generated power, p1 (supplier node) is equal to the negative sum of (p2 + p3 + p4).
    \item \textbf{g1—g4}: Price elasticity coefficient of each network participant, with actual values ranging from 0.05 to 1 (g1 corresponds to the supplier node, g2 to g4 correspond to consumer nodes; g is gamma).
\end{itemize}

The description of the feature dataset characteristics is shown in Table 1, which includes all discrete and continuous outputs used for regression and classification through feature selection using feature engineering. The results of the P-value hypothesis test showed the importance of features related to each network's reaction time and price elasticity, as shown in Table 2.

\begin{table}[h]
\centering
\caption{Summary of feature datasets for big data analysis of smart grid systems}
\label{table:1}
\begin{tabular}{|c|c|c|c|}
\hline
\textbf{Feature} & \textbf{Count} & \textbf{Min Value} & \textbf{Max Value} \\
\hline
tau1 & 60000 & 0.5 s & 10.0 s \\
tau2 & 60000 & 0.5 s & 10.0 s \\
tau3 & 60000 & 0.5 s & 10.0 s \\
tau4 & 60000 & 0.5 s & 10.0 s \\
p1 & 60000 & 1.5 & 6 \\
p2 & 60000 & -2.0 & -0.5 \\
p3 & 60000 & -2.0 & -0.5 \\
p4 & 60000 & -2.0 & -0.5 \\
g1 & 60000 & 0.05 & 1 \\
g2 & 60000 & 0.05 & 1 \\
g3 & 60000 & 0.05 & 1 \\
g4 & 60000 & 0.05 & 1 \\
Stab & 60000 & -0.08 & 0.11 \\
Staff & 60000 & Unstable & Stable \\
\hline
\end{tabular}
\end{table}

\begin{table}[h]
\centering
\caption{Application of feature engineering and selection of input features for classification}
\label{table:2}
\begin{tabular}{|c|c|c|c|}
\hline
\textbf{Feature} & \textbf{P-value} & \textbf{Correlation} & \textbf{Importance} \\
\hline
tau1 & 0.0001 & -0.2181 & Important \\
tau2 & 0.0001 & -0.2208 & Important \\
tau3 & 0.0001 & -0.2249 & Important \\
tau4 & 0.0001 & 0.1884 & Important \\
p1 & 0.9700 & 0.0024 & Not Important \\
p2 & 0.9800 & 0.0016 & Not Important \\
p3 & 0.9100 & 0.0007 & Not Important \\
p4 & 0.9500 & -0.0048 & Not Important \\
g1 & 0.0001 & 0.3920 & Important \\
g2 & 0.0001 & -0.4202 & Important \\
g3 & 0.0001 & -0.4184 & Important \\
g4 & 0.0001 & 0.1822 & Important \\
\hline
\end{tabular}
\end{table}

\subsection{Model fitting and prediction accuracy analysis}

This paper provides 3 machine learning models: decision tree, random forest and deep learning.

Decision trees involve the use of nodes and branches to enable the model to learn and distinguish accuracy. The difference between decision trees and random forests lies in the complexity of nodes and branches. The random forest algorithm uses different decision trees at each node to provide optimal explanations through combining decision trees. For classification problems, the Gini index is used to identify nodes and branches in decision trees, as shown in Equation (4):

\begin{equation}
Gini = 1 - \sum_{i=1}^{c} p_i^2 \quad 
\end{equation}

where \( c \) is the number of categories, and \( p_i \) is the relative frequency of each category.

Deep learning is a subfield of machine learning that leverages artificial neural network methods to enhance model learning and data integration. The deep learning process is given in Equation (5) and is used to create a prediction model for the distributed smart grid control system:

\begin{equation}
z = \sum_{i=1}^{n} w_i x_i + b \quad 
\end{equation}

where \( z \) is the predicted output value of the deep learning model, \( x_i \) is the variable, \( w_i \) is the weight of each variable, and \( b \) is the bias.

The results from the linear regression model are shown in Fig. 4. It has been demonstrated that when the amount of training data is sufficient, the accuracy of the implemented model can reach over 95\%.
\begin{figure}[h]
\centering
\includegraphics[width=0.5\textwidth]{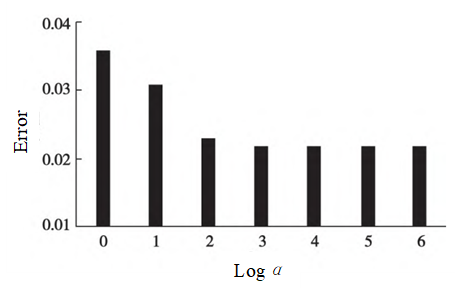}
\caption{Penalty linear regression diagram for power grid stability of power generation enterprises}
\label{fig:stability_histogram}
\end{figure}

\section{Conclusion}\label{Sec:Conclusion}

This paper establishes an implementation framework for big data analysis in integrated smart energy project document management and proposes a method to optimize the efficiency of integrated smart energy project document management through machine learning. Big data analysis and machine learning have great potential for integrating various documents generated in the comprehensive energy project management of power generation enterprises, promoting digital and intelligent transformation in the field of document management in the energy industry. By leveraging these cutting-edge technologies, power generation enterprises can gain valuable insights into their operations, identify inefficiencies and areas for improvement, and make data-driven decisions to enhance competitiveness and profitability.
    
The framework proposed in this paper can help power generation enterprises optimize their energy production processes, reduce emissions and costs, and improve overall efficiency and sustainability. However, this integration also presents challenges such as data privacy, data quality, and the need for skilled professionals to operate these complex systems. Therefore, power generation enterprises should implement these technologies in a strategic and responsible manner to fully realize the benefits of this innovative energy management approach.

\bibliographystyle{IEEEtran}
\bibliography{ref.bib} 

\end{document}